\documentclass[12pt]{iopart}

\usepackage{amsmath}  
\usepackage[british]{babel}
\usepackage[utf8]{inputenc}
\usepackage{graphicx}
\usepackage[per-mode=symbol]{siunitx}
\usepackage{hyperref}
\usepackage{textcomp}

\begin{document}

\title[Thermodynamics of Trapped Photon Gases]{Thermodynamics of Trapped Photon Gases\\at Dimensional Crossover from 2D to 1D}

\author{Enrico Stein}
\ead{estein@rhrk.uni-kl.de}
\address{Department of Physics and Research Center OPTIMAS, Technische Universität Kaiserslautern, Erwin-Schrödinger Straße 46, 67663 Kaiserslautern, Germany}

\author{Axel Pelster}
\ead{axel.pelster@physik.uni-kl.de}
\address{Department of Physics and Research Center OPTIMAS, Technische Universität Kaiserslautern, Erwin-Schrödinger Straße 46, 67663 Kaiserslautern, Germany}

\vspace{10pt}
\begin{indented}
\item[]\today
\end{indented}

\begin{abstract}
Photon Bose-Einstein condensates are characterised by a quite weak interaction, so they behave nearly as an ideal Bose gas. Moreover, since the current experiments are conducted in a microcavity, the longitudinal motion is frozen out and the photon gas represents effectively a two-dimensional trapped gas of massive bosons. In this paper we focus on a harmonically confined ideal Bose gas in two dimensions, where the anisotropy of the confinement allows for a dimensional crossover. If the confinement in one direction is strong enough so that this squeezed direction is frozen out, then only one degree of freedom survives and the system can be considered to be quasi-one dimensional. In view of an experimental set-up we work out analytically the thermodynamic properties for such a system with a finite number of photons. In particular, we focus on examining the dimensional information which is contained in the respective thermodynamic quantities.
\end{abstract}

%
\vspace{2pc}
\noindent{\it Keywords}: Photon Bose--Einstein Condensate, Thermodynamics, Dimensional Crossover\\
%
\noindent{\submitto{\NJP}}
%
\maketitle

\section{Introduction}
\label{sec:intro}
The question of Bose-Einstein condensation in lower dimensions got already tackled quite early in the post-war era of physics. Soon it was found out that in the case of lower dimensional systems without trapping potential no long-range order can emerge \cite{PhysRevLett.17.1133, Hohenberg1967} and, thus, no Bose-Einstein condensation in such systems is possible. Later on in the early 1990s but prior to the experimental realisation of Bose-Einstein condensates (BEC), the authors of Ref.~\cite{Bagnato1991} worked out that, with the aid of an external trapping potential, the excited states of lower-dimensional ideal Bose gases can saturate, meaning that Bose-Einstein condensation is possible, whereas in three spatial dimensions this is also possible for the non-trapped case. In the thermodynamic limit they showed for a trapping potential, which is stronger confining than a box in the sense of a monomial spatial dependence $\sim x^\alpha$, that a condensation in 2D can occur, whilst in a 1D setting a potential more confining than a quadratic potential is necessary. Soon after this a full quantum mechanical follow-up study \cite{Ketterle1996} revealed that for the harmonically trapped 1D-Bose gas with a finite number of particles BEC is possible. Furthermore, Ref.~\cite{Yukalov2005} generalised the semiclassical ansatz from Ref.~\cite{Bagnato1991} and showed, that these improved results agree with the corresponding finite-size results from \cite{Ketterle1996}. After the experimental realisation of BECs \cite{Anderson1995, Davis1995} naturally the question came up, how to achieve systems with an effective dimension lower than three.\\
In the experimental work of Ref.~\cite{Gorlitz2001a} the question of the effective dimensionality of the system is reduced to a comparison of different length scales, which are in a 3D axially symmetrically trapped Bose gas the in-plane radius, the axial width, the scattering length, and the healing length. A system is effectively 2D, if the healing length is larger than the axial width, and effectively 1D, if the healing length is larger than the in-plane radius with the axial width being still larger than the healing length. As the healing length is inversely proportional to the square root of both the density and the scattering length, one can control the effective dimension either by changing the density, as is done in Ref.~\cite{Gorlitz2001a}, or by changing the interaction strength itself via a Feshbach resonance \cite{RevModPhys.82.1225}. Nowadays, harmonically trapped 1D and 2D condensates can not only be found in atomic systems \cite{Lausch2018} but also in condensates of light created in 1D-fiber cavities \cite{Weill2019}.\\
Another possibility of tuning the effective dimension of a system is to modify the kinetic energy in a certain direction. This can be achieved in lattice systems by changing the hopping matrix amplitude, e.g.~by changing the lattice depth in a certain direction. One possible experimental realisation is to use coupled 2D BECs in order to perform a crossover from 2D to 3D \cite{Cennini2006}. A more recent experiment \cite{PhysRevLett.113.215301}, which is described by the theoretical works \cite{Cazalilla2006, PhysRevA.95.043610}, consists of 2D arrays of coupled 1D Bose gases. Decreasing the 2D lattice depth yields an increase of the hopping amplitude and thus gives rise to a dimensional crossover to higher dimensions.\\
In the following, however, we dedicate our discussion for the sake of concreteness on photon BECs as realised in Ref.~\cite{Klaers2010a, Walker2018, Greveling2018}. As these kinds of experiments are conducted in a microcavity, the direction along the optical axis is already frozen out, since it corresponds to a standing wave along this very direction. Therefore, these systems are intrinsically two-dimensional. On the one hand experiments have found an effective photon-photon interaction \cite{Klaers2010a, keijsers2020}, which is explained by a thermo-optic effect, and also theoretical investigations \cite{vanderWurff2014, Alaeian2017, Stein2019} have revealed the influence of this kind of interactions on the photon BEC. Contrarily to that experimental measurements have also figured out that the thermodynamical behaviour of this system is not affected by this kind of interaction \cite{Damm2016}. In this respect the photon BEC can be seen as a realisation of an ideal Bose gas. Thus, the question remains how to define and how to determine the effective dimension of the gas when changing from an isotropic harmonic confinement to a highly anisotropic confinement giving rise for a dimensional crossover from a two-dimensional gas to a quasi-1D gas. To this end, we work out how the thermodynamic quantities change as a function of the trap-aspect ratio. In particular, we carefully analyse not only the thermodynamic limit but also the respective finite-size corrections similarly to a corresponding seminal study in 3D \cite{Grossmann1995}, whose predictions where experimentally confirmed despite of systematic measurement errors for thermodynamic quantities in \cite{Gerbier2004}. As we find in this work that these finite-size corrections increase by lowering the dimension of the system, we expect that they can also be confirmed in the dimensional crossover of photon BECs. Furthermore, in this setting the detection of finite-size effects is even more straight-forward than for atomic BECs, as these effects are not masked by interaction effects. As so far photon BEC experiments have only been performed in an isotropic setup, this theoretical paper paves the way towards future experiments with strongly anisotropic harmonic trapping potentials. Such potentials can be achieved, for instance, by ellipsoidally grinding the mirrors or by heat induced mirror surface delamination \cite{Kurtscheid2019, Kurtscheid2020} which allows, however, only for traps with small anisotropies due to the limited resolution. Thus, in view of achieving stronger anisotropies it is more promising to use direct laser writing \cite{Maruo97, Deubel2004, Hohmann2015} or focused ion beam milling \cite{Walker2021} as a microstructuring technique, as it is then possible to create potential landscapes with spatial variations of the order of the wavelength of the photons.\\
This paper is organised as follows. Section \ref{Sec:Potential} introduces the setting and provides an analytical expression for the thermodynamic potential of an ideal Bose gas at the dimensional crossover from 2D to 1D. Equipped with this, Sec.~\ref{Sec:Number} specialises to the photon gas and derives expressions for the critical particle number as well as for the condensate fraction. Afterwards, the specific heat of the photon gas is discussed in Sec.~\ref{Sec:Spec}, which is finally used to define the effective dimension of the system in Sec.~\ref{Sec:Phase_Dim}.

\section{Grand-Canonical Potential \label{Sec:Potential}}
At first we analyse the thermodynamic properties of an ideal Bose gas at the dimensional crossover between 2D and 1D. To this end we consider a two-dimensional harmonic trap for bosons, where the trapping frequency in $y$-direction can be altered. Thus, with the quantum numbers $j,n$ in the respective dimensions the energy levels are given by:
\begin{align}\label{Eq:Energy}
	E_{jn}(\lambda) = \hbar\Omega\left(j+\lambda n + \frac{1+\lambda}{2}\right),
\end{align}
where $\Omega$ stands for the trapping frequency in $x$-direction and $\lambda = \Omega_y/\Omega$ denotes the trap-aspect ratio.
We remark, that for a isotropic 2D oscillator, which we will call the 2D case in the following, we have $\lambda=1$, whereas the one-dimensional case is approached in the limit $\lambda \rightarrow\infty$. In this paper we always fix the trapping frequency $\Omega$ in $x$-direction and increase the trap-aspect ratio $\lambda$, corresponding to a squeezing in the $y$-direction. Intuitively, the gas can already be considered to be effectively one dimensional, if the energy spacing $\lambda\hbar\Omega$ in $y$-direction is larger than the thermal energy $k_\text{B}T$, which leads for the trap-aspect ratio to the condition
\begin{align}\label{Eq:Crit_1D}
	\lambda > \lambda_\text{1D}.
\end{align}
Here we define the effective one-dimensional trap-aspect ratio $\lambda_\text{1D}$, which depends on the temperature $T$ of the system and comprises the Boltzmann constant $k_\text{B}$ as well as the reduced Planck constant $\hbar$: 
\begin{align}\label{Eq:l1D}
	\lambda_\text{1D} = \frac{k_\text{B}T}{\hbar\Omega}.	
\end{align}
Again we point out that regarding the experimental findings in \cite{Damm2016} it is a very good approximation to neglect the effective photon-photon interaction for discussing the thermodynamic properties of the photon BECs.\\
Taking into account the energy levels \eqref{Eq:Energy}, we have with the chemical potential $\mu$, the inverse temperature $\beta=1/(k_\text B T)$, and the degeneracy $g$, which takes for photons the two polarisation degrees of freedom into account, the grand-canonical potential \cite{Schwabl2006}
\begin{align}\label{Eq:potential}
	\Pi = -\frac{g}{\beta}\sum_{j,n=0}^\infty \sum_{k=1}^\infty \frac{e^{-\beta\left[E_{jn}(\lambda)-\mu\right]k}}{k},
\end{align}
where we have used the series representation of the logarithm $\ln(1+x) = -\sum_{k=1}^\infty (-x)^k/k$.
Performing the sum over the energy levels in the non-squeezed direction, which are labelled by $j$, allows us to write the potential $\Pi$ in the form of a dimensional expansion
\begin{align}\label{Eq:Gen_pontential}
	\Pi = \Pi_\text{1D} + \Delta\Pi(\lambda).	
\end{align}
Here the one-dimensional grand-canonical potential reads
\begin{align}\label{Eq:Pi1D}
	\Pi_\text{1D} = -g\frac{\hbar\Omega}{b}I\left(\tilde\mu, b, -1\right),
\end{align}
where we introduced the dimensionless variables $b = \beta\hbar\Omega$ and $\tilde\mu = (1+\lambda)/2-\mu/(\hbar\Omega)$ as well as the auxiliary function
\begin{align}\label{Def:I}
	I(a, b, l) = \sum_{k=1}^\infty k^l \frac{e^{-abk}}{1-e^{-bk}},
\end{align}
cf.~Ref.~\cite{Klunder2009}. The correction $\Delta\Pi(\lambda)$ to the 1D potential, which takes the second dimension into account, depends on the trap-aspect ratio $\lambda$ via
\begin{align}\label{Eq:DPi}
	\Delta\Pi(\lambda) = -g\frac{\hbar\Omega}{b}\sum_{n=1}^\infty I\left(\tilde\mu + \lambda n, b, -1\right).
\end{align}
The auxiliary function $I(a, b, -1)$ appearing in \eqref{Eq:Pi1D} and \eqref{Eq:DPi} is determined in \ref{App:Regularisation} in the form \eqref{Eq:IM1} and \eqref{Eq:AppS}, respectively. Therefore, we find for the total grand-canonical potential analytically:
\begin{align}\label{Eq:Pia}
	\Pi=&-g\frac{\hbar\Omega}{b}\left\{f(\tilde\mu) +\frac{1}{b}\zeta_2\left(e^{-\tilde\mu b}\right)+\frac{1}{2}\zeta_1\left(e^{-\tilde\mu b}\right)+\frac{b}{12}\zeta_{0}\left(e^{-\tilde\mu b}\right)\right\}\nonumber\\
	&-g\frac{\hbar\Omega}{b}\left\{\frac{1}{\lambda b^2} \zeta_3\left(e^{-(\tilde\mu+\lambda)b}\right) + \frac{1}{2\lambda b}\zeta_2\left(e^{-(\tilde\mu+\lambda)b}\right)+ \frac{1}{12\lambda}\zeta_1\left(e^{-(\tilde\mu+\lambda)b}\right) + \frac{1}{\lambda} \int_{\tilde\mu+\lambda}^\infty dy~f(y) \right.\nonumber\\
    &\left.+ \frac{1}{2}\left[f(\tilde\mu+\lambda) +\frac{1}{b}\zeta_2\left(e^{-(\tilde\mu+\lambda)b}\right)+\frac{1}{2}\zeta_1\left(e^{-(\tilde\mu+\lambda)b}\right)+\frac{b}{12}\zeta_{0}\left(e^{-(\tilde\mu+\lambda)b}\right)\right]\right\}+\dots.
\end{align}
Here, $\zeta_l(x)=\sum_{k=1}^\infty x^k/k^l$ denotes the polylogarithm \cite{Gradshteyn2007} and $f(\tilde\mu)$ is defined in Eq.~\eqref{Eq:DefI}. The dots indicate here and in the following terms of order $b^2$ and higher. We remark, that the one-dimensional limit is given by $\lambda\rightarrow\infty$, which corresponds to the vanishing of the last two lines in Eq.~\eqref{Eq:Pia}.\\
In the following we discuss the thermodynamic consequences of the grand-canonical potential \eqref{Eq:Pia} for a general ideal Bose gas, but for illustrating the functional dependencies of the thermodynamic quantities we specialise these general results to the photon BEC experiments in Bonn \cite{Klaers2010a, Damm2016}. There, we have to take into account the two polarisational degrees of freedom of the photons resulting in the degeneracy $g=2$. For typical values, i.e.~$T_0=\SI{300}{\kelvin}$ and $\Omega = 2\pi\times\SI{40}{\giga\hertz}$, the system can be considered to be effectively one dimensional, if the trap-aspect ratio fulfills condition \eqref{Eq:Crit_1D} with $\lambda_\text{1D}\approx 156$. Moreover, since the photon BEC experiment is performed at room temperature $T_0$ \cite{Klaers2010a, Damm2016}, the approximation of small $b$ is well fulfilled, as we have then $b_0\equiv \hbar\Omega/k_\text B T_0 \approx \num{6e-3}$. Note, that the same order of magnitude is also obtained for the atomic BEC case, when taking the experimental parameters from \cite{Davis1995}.

\section{Particle Number \label{Sec:Number}}
By calculating the derivative $N=-\partial \Pi/\partial \mu$, we find from the potential \eqref{Eq:Pia} for the total particle number
\begin{align}\label{Eq:N}
	N =&g\left\{- \frac{1}{b} f'(\tilde\mu) + \frac{1}{b}\zeta_1\left(e^{-\tilde\mu b}\right) + \frac{1}{2}\zeta_0\left(e^{-\tilde\mu b}\right) + \frac{b}{12}\zeta_{-1}\left(e^{-\tilde\mu b}\right)\right.\\
	& +\frac{1}{\lambda b^2}\zeta_2\left(e^{-(\tilde\mu+\lambda)b}\right) + \frac{1}{2\lambda b}\zeta_1\left(e^{-(\tilde\mu+\lambda)b}\right) + \frac{1}{12\lambda}\zeta_0\left(e^{-(\tilde\mu+\lambda)b}\right) + \frac{1}{\lambda b}f(\tilde\mu+\lambda)\nonumber\\
	&\left.+\frac{1}{2}\left[-\frac{1}{b}f'(\tilde\mu+\lambda) + \frac{1}{b}\zeta_1\left(e^{-(\tilde\mu+\lambda)b}\right) + \frac{1}{2}\zeta_0\left(e^{-(\tilde\mu+\lambda)b}\right) + \frac{b}{12}\zeta_{-1}\left(e^{-(\tilde\mu+\lambda)b}\right)\right]\right\}+\dots.\nonumber
\end{align}
This explicit expression allows to determine the critical particle number, the critical temperature and, likewise, the condensate fraction as will be derived in section \ref{ssec:Nc}, \ref{ssec:Tc} and \ref{ssec:condfrac}, respectively. Already here, we mention that we will indeed find a critical particle number and critical temperature in the 1D case in accordance with Refs.~\cite{Ketterle1996, Yukalov2005, Klunder2009, Haugset1997}. For further detail, we refer to the discussion at the beginning of section \ref{ssec:Tc}.  

\subsection{Critical Particle Number}
\label{ssec:Nc}
\begin{figure}
    \centering
    \includegraphics[width=\linewidth]{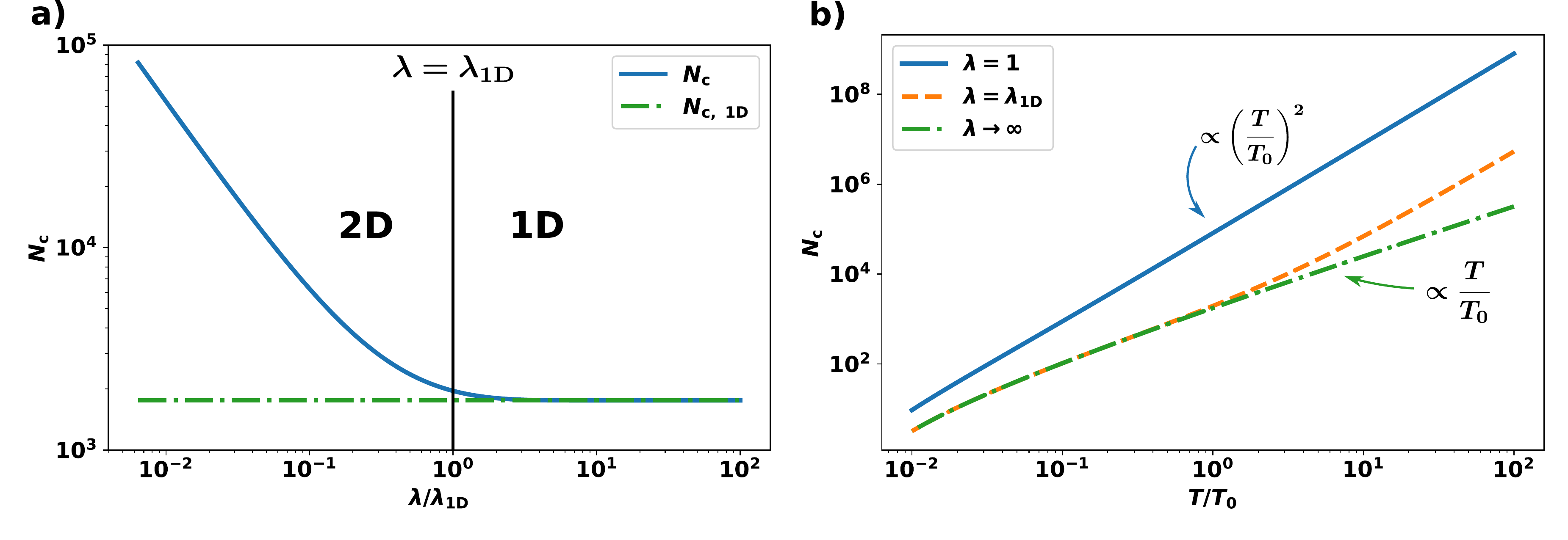}
    \caption{\textbf{a)} Critical particle number \eqref{Eq:Nc} at room temperature $T_0$ for varying trap-aspect ratio (blue/solid line). The green (dashed dotted) line illustrates the 1D limit \eqref{Eq:Nc1D}.
    \textbf{b)} Critical particle number \eqref{Eq:Nc} for different trap-aspect ratios $\lambda$ as a function of the temperature $T$ normalised to the room temperature $T_0$. The blue (solid) line represents the isotropic 2D case, i.e.~$\lambda=1$, the orange (dashed) line is for $\lambda=k_\text B T_0/\hbar\Omega\equiv\lambda_\text{1D}$, and the green (dash-dotted) depicts the 1D limit, i.e.~$\lambda\rightarrow\infty$.}
    \label{fig:Nc}
\end{figure}

In order to calculate the critical particle number, we consider the deep condensate limit $\tilde\mu\rightarrow0$. We remark, that this limit corresponds to the order parameter approach, as worked out in Ref.~\cite{Klunder2009}, where the ground-state particle number is used as an order parameter for the BEC phase transition and only the excited states are treated in a thermodynamic way. This approach corresponds to describing the Bose gas in the thermodynamic limit. In the present work, however, we treat all states, including the ground state, thermodynamically as this description is closer to the experimental situation, where the system is finite.
With this the particle number \eqref{Eq:N} can be written in the form
\begin{align}\label{Eq:NBEC}
    N \approx N_0 + N_\text c,
\end{align}
with the ground-state particle number 
\begin{align}
    N_0 = \frac{g}{e^{\tilde\mu b}-1},    
\end{align}
 which acquires in the limit $\tilde\mu\rightarrow0$ the form $N_0 \approx g/(\tilde\mu b)$, and the critical particle number
\begin{align}\label{Eq:Nc}     
	N_\text{c} =&g\frac{\gamma-\ln(b)}{b}&\nonumber\\
	&+g\left\{\frac{1}{\lambda b^2}\zeta_2\left(e^{-\lambda b}\right) + \frac{1}{2\lambda b}\zeta_1\left(e^{-\lambda b}\right) + \frac{1}{12\lambda}\zeta_0\left(e^{-\lambda b}\right) + \frac{1}{\lambda b}f(\lambda)\right.\nonumber\\
	&\left.+\frac{1}{2}\left[-\frac{1}{b}f'(\lambda) + \frac{1}{b}\zeta_1\left(e^{-\lambda b}\right) + \frac{1}{2}\zeta_0\left(e^{-\lambda b}\right) + \frac{b}{12}\zeta_{-1}\left(e^{-\lambda b}\right)\right]\right\}+\mathcal{O}\left((\tilde\mu b)^0\right),
\end{align}
with the Euler-Mascheroni constant $\gamma\approx0.577$.
Note that, due to the limit process involved, this result is only accurate up to order $\mathcal{O}\left((\tilde\mu b)^0\right)$, but it is still accurate to all orders of $\lambda b$. 
Moreover, we note the same structure as for the grand-canonical potential in Eq.~\eqref{Eq:Gen_pontential}, namely the bare one-dimensional quantity in the first line gets modified by the terms in the other two lines, which depend on the trap-aspect ratio and describe the influence of the second dimension, see Fig.~\ref{fig:Nc} a).
Note that the first line in \eqref{Eq:Nc} follows from the first line in Eq.~\eqref{Eq:N} by applying the Robinson formula \cite{Robinson1951},
\begin{align}\label{Eq:Robinson}
    \zeta_{l}\left(e^{-a}\right)=\frac{(-a)^{l-1}}{(l-1) !}\left\{\sum_{k=1}^{l-1} \frac{1}{k}-\ln a\right\}+\sum_{k=0 \atop k \neq l-1}^{\infty} \frac{(-a)^{k}}{k !} \zeta(l-k),
\end{align}
where $\zeta(l)$ denotes the Riemann-$\zeta$ function and $a>0$, in order to expand the occurring polylogarithms with positive integer index $l$ and by using for the corresponding polylogarithms with negative index the representation
\begin{align}
    \zeta_{-l}\left(e^{-a}\right)=\frac{1}{\left(1-e^{-a}\right)^{l+1}}\sum_{k=0}^{l-1} \left\langle l\atop k\right\rangle e^{-(l-k)a}
\end{align}
with the Eulerian numbers $\left\langle l\atop k\right\rangle$ \cite{NIST:DLMF}.\\
From \eqref{Eq:Nc} we find for the 1D critical particle number
\begin{align}\label{Eq:Nc1D}
	N_\text{c, 1D} = g\frac{k_\text B T}{\hbar\Omega} \left[\gamma-\ln\left(\frac{\hbar\Omega}{k_\text B T}\right)\right].
\end{align}
Near two dimensions, meaning small anisotropy $\lambda\ll1/b$, the critical particle number \eqref{Eq:Nc} reads
\begin{align}\label{Eq:Nc2D1}
	N_\text{c, $\approx$2D} = N_\text{c, 1D} + g\left\{\frac{\zeta(2)}{\lambda (\hbar\Omega\beta)^2}+\frac{1}{2\hbar\Omega\beta}\left[\ln(\hbar\Omega\beta)+\frac{\ln\Gamma(\lambda)-\ln(2\pi \hbar\Omega\beta)}{\lambda}-\psi_0(\lambda)\right]\right\},	
\end{align}
where $\Gamma(x) = \int_0^\infty dt~t^{x-1}e^{-t}$ denotes the $\Gamma$-function and $\psi_0(x) = \partial_x\ln\Gamma(x)$ is the digamma function. In two dimensions, i.e.~$\lambda=1$, this reduces to
\begin{align}\label{Eq:Nc2D}
	N_\text{c, 2D} = N_\text{c, 1D} + g\left\{\zeta(2)\left(\frac{k_\text B T}{\hbar\Omega}\right)^2+\frac{k_\text B T}{2\hbar\Omega}\left[\gamma-\ln(2\pi)\right]\right\}.
\end{align}
At first, a comparison with Ref.~\cite{Klunder2009} shows that Eq.~\eqref{Eq:Nc1D} is exact, whereas the corresponding expression for the two-dimensional critical particle number \eqref{Eq:Nc2D} contains the last term in addition. This difference is solely due to our approach, where we calculate at first the one-dimensional quantities and approximate afterwards the corresponding two-dimensional ones. As the leading order of the relative deviation of our result \eqref{Eq:Nc2D} compared to the corresponding one in Refs.~\cite{Klunder2009, Xie2018} is of the order of the magnitude of $b$ itself, the difference for the experimental parameter regime, i.e.~room temperature $T_0 = \SI{300}{\K}$ and $\Omega=2\pi\times\SI{40}{\giga\Hz}$, is of the order $b_0\approx\num{6e-3}$ and, thus, negligible for all practical purposes.\\ 
Already here, we also encounter a qualitative difference between the two special cases of dimensions. In 1D the critical particle number \eqref{Eq:Nc1D} depends linearly on the temperature, apart from the logarithmic term, whereas in 2D the leading order in Eq.~\eqref{Eq:Nc2D} is quadratic in the temperature.
In Fig.~\ref{fig:Nc} b) we plot the critical particle number \eqref{Eq:Nc} as a function of the temperature for different trap-aspect ratios $\lambda$. Neither in the 2D case, $\lambda=1$, nor in the 1D case, which amounts to the limit $\lambda\rightarrow\infty$, the functional dependence of the critical particle number on the temperature changes. However, we note the different exponents one and two in accordance with \eqref{Eq:Nc1D} and \eqref{Eq:Nc2D}, which can be interpreted as a sign for the corresponding dimension. For an intermediate trap-aspect ratio of $\lambda=k_\text B T_0/(\hbar\Omega)\equiv \lambda_\text{1D}$ the temperature dependence changes qualitatively for $T\approx T_0$. For smaller temperatures the curve coincides with the 1D curve, while for larger temperatures the orange curve gets parallel to the 2D curve. This means, that in the former case the system behaves effectively one dimensional, whereas in the latter case the system reveals a two-dimensional behaviour. This observation completely agrees with the criterion \eqref{Eq:Crit_1D} for quasi one-dimensionality.

\subsection{Critical Temperature}
\label{ssec:Tc}
\begin{figure}
	\centering
	\includegraphics[width=.75\linewidth]{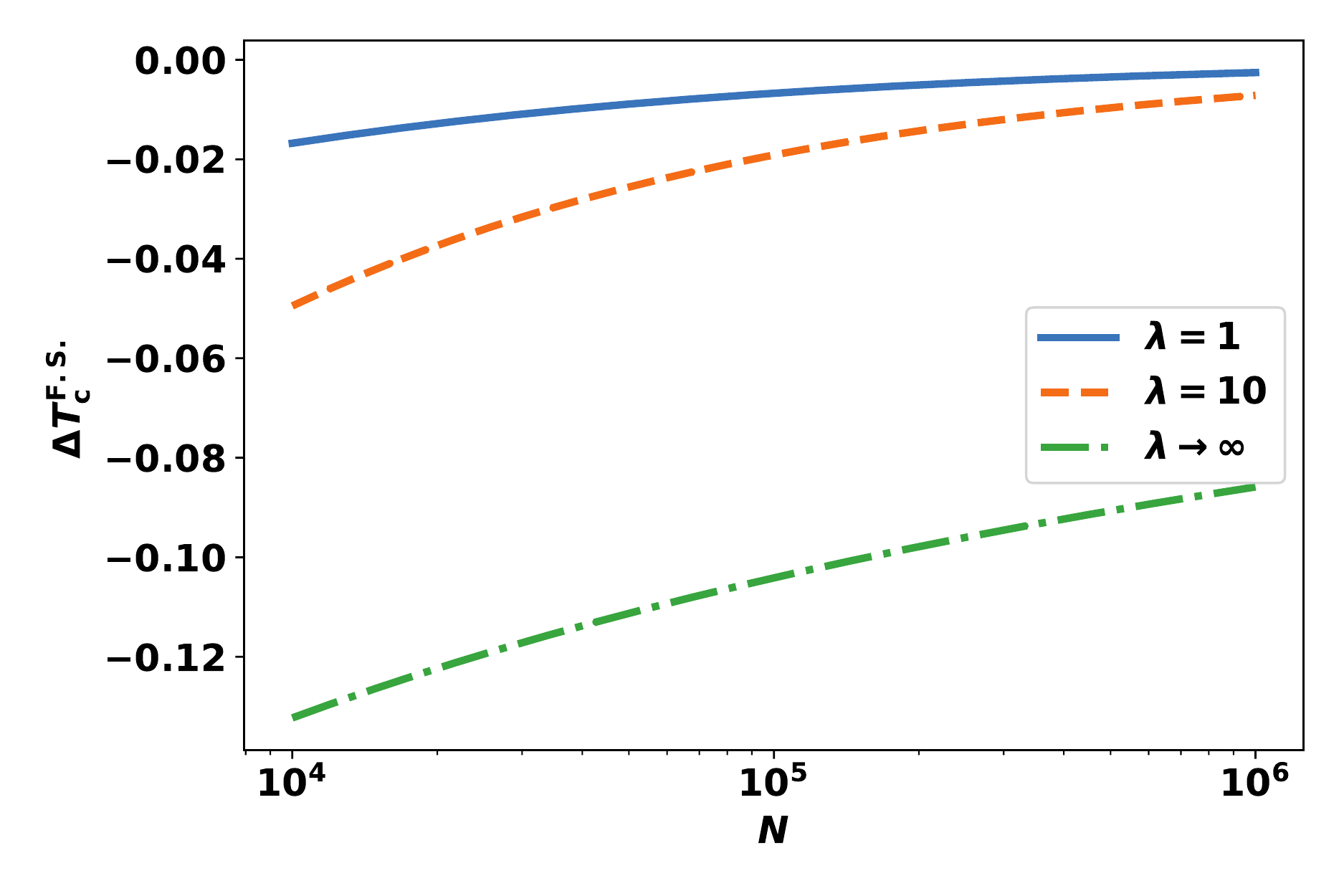}
	\caption{Finite-size corrections \eqref{Eq:FS} to the critical temperature in 2D (blue), for finite trap-aspect ratio $\lambda=10$ (orange, dashed) and in the 1D case $\lambda\rightarrow\infty$ (green, dashed-dotted) according to \eqref{Eq:bc2DFS}, \eqref{Eq:bc2D1FS} and \eqref{Eq:bc1DFS}, respectively.}
	\label{Fig:finite_size}
\end{figure}

Finally, we also solve the critical particle number $N(T_\text c)$ for the critical temperature $T_\text c$ in the respective dimension. In 1D, we obtain from directly inverting Eq.~\eqref{Eq:Nc1D} the implicit equation
\begin{align}\label{Eq:bc1D}
    T_\text{c, 1D} = \frac{\hbar\Omega}{gk_\text B}\frac{N}{\gamma-\ln (\hbar\Omega/k_\text B T_\text{c, 1D})}.
\end{align}
In \cite{Bagnato1991} it is derived, that no Bose-Einstein condensation is possible for a harmonic confining potential in one spatial dimension, as the critical temperature tends to zero in this limit. In contrast to that, we find in \eqref{Eq:bc1D} a finite critical temperature and, thus, the possibility for a Bose-Einstein condensate. The difference between the approach in \cite{Bagnato1991} and our approach is, that the former work is performed entirely in the thermodynamic limit, whereas we always assume a finite system size. A further difference is that the approach in \cite{Bagnato1991} relies on the density of states, whereas we directly evaluate the appearing sum in \eqref{Eq:potential}. Therefore, on a mathematical basis the divergent value $\zeta(1)$, which is obtained by \cite{Bagnato1991} in the limit of a harmonic trapping potential, is in our calculation resolved by the logarithm appearing in \eqref{Eq:bc1D} due to our finite-size ansatz. Furthermore, we emphasise that our result is also obtained in \cite{Klunder2009}, where an order-parameter approach has been used, and similar results, which are also based on a full thermodynamic approach, but with a less systematic application of the Euler-Maclaurin formula, are found in \cite{Haugset1997}. Finally, we remark that due to this finite-size behaviour, we will refer to the one-dimensional condensed phase still as BEC in accordance with Refs.~\cite{Ketterle1996, Yukalov2005, Gorlitz2001a}.\\ 
In order to figure out an approximate solution to the transcendental equation \eqref{Eq:bc1D} we iterate \eqref{Eq:bc1D} once and neglect the further logarithmic dependencies. For large photon numbers $N$ we have then a leading term of the critical temperature in 1D, which is already determined by an inverse logarithmic dependenc\
\begin{align}\label{Eq:bc1DLT}
	T_\text{c, 1D}^\text{L.T.} \approx \frac{\hbar\Omega}{gk_\text b}\frac{N}{\ln(N/g)}.	
\end{align}
We remark, that this critical temperature in 1D can be derived by a generalisation of the semiclassical approach used in \cite{Bagnato1991} as has been shown in \cite{Yukalov2005} and is backed up experimentally, e.g.~\cite{Lausch2018, Weill2019}. However, we emphasise here, that the thermodynamic limit relies on a finite $T_\text c$. Thus, according to the numerator of \eqref{Eq:bc1DLT} this implies $\Omega N=\text{const}$. Therefore, from the denominator of \eqref{Eq:bc1DLT} we have then $T_\text c\rightarrow0$ in the limit $N\rightarrow\infty$. With this we rederive the results of \cite{Bagnato1991} from our finite-size considerations in accordance with \cite{Yukalov2005}. Consequently, in the thermodynamic limit, indeed, no Bose-Einstein condensation is possible at any finite temoerature. However, in the realistic experimental settings, where the particle number $N$ may be large but finite, Bose-Einstein condensation can always be observed in a one-dimensional harmonic trap. \\
The finite-size corrections we define by 
\begin{align}\label{Eq:FS}
	\Delta T_\text{c, $\bullet$}^\text{F.S.} = \frac{T_\text{c, $\bullet$}-T_\text{c, $\bullet$}^\text{L.T.}}{T_\text{c, $\bullet$}^\text{L.T.}},
\end{align}
where the bullet stands for the corresponding dimension. In 1D we have for the finite-size corrections \eqref{Eq:FS}
\begin{align}\label{Eq:bc1DFS}
	\Delta T_\text{c, 1D}^\text{F.S.} = \frac{\ln(\gamma)-\gamma}{\ln(N/g)},
\end{align}
which is also determined by a logarithm.\\
In nearly two dimensions we find by iterating Eq.~\eqref{Eq:Nc2D1} once for $b\lambda\ll1$ as leading term
\begin{align}\label{Eq:bc2D1LT}
	T_\text{c, $\approx$2D}^\text{L.T.} = \frac{\hbar\Omega}{k_\text B} \sqrt{\frac{\lambda N}{g\zeta(2)}},
\end{align}
which reduces in 2D to
\begin{align}\label{Eq:bc2DLT}
	T_\text{c, 2D}^\text{L.T.} =\frac{\hbar\Omega}{k_\text B} \sqrt{\frac{N}{g\zeta(2)}}.
\end{align}
The finite-size corrections \eqref{Eq:FS} near 2D are given by
\begin{align}\label{Eq:bc2D1FS}
	\Delta T_\text{c, $\approx$2D}^\text{F.S.} = - \frac{\lambda}{4\zeta(2)}\sqrt{\frac{g\zeta(2)}{\lambda N}}&\left[ 2\gamma-\psi_0(\lambda)+\frac{\ln\Gamma(\lambda)-\ln\left(2\pi\sqrt{g\zeta(2)/(\lambda N)}\right)}{\lambda}\right.\nonumber\\
 &\left.-\frac{1}{2}\ln\left(\frac{g\zeta(2)}{\lambda N}\right)\right]	
\end{align}
simplifying in 2D to the form
\begin{align}\label{Eq:bc2DFS}
	\Delta T_\text{c, 2D}^\text{F.S.} =- \frac{1}{4\zeta(2)}\sqrt{\frac{g\zeta(2)}{N}}\left[3\gamma-\ln\left(2\pi\frac{g\zeta(2)}{N}\right)\right].
\end{align}
Comparing Eqs.~\eqref{Eq:bc1DLT} and \eqref{Eq:bc2DLT} we note the following two differences between the two limiting cases of the dimension. At first, in 1D the total particle number contributes with the exponent one to the critical temperature, whereas in 2D it appears with a square root. A second difference between 1D and 2D is the occurrence of the logarithm. In 1D the logarithm shows up already in the leading term \eqref{Eq:bc1DLT} of the critical temperature, while in 2D the logarithm determines the first finite-size correction \eqref{Eq:bc2DFS}.\\
In Fig.~\ref{Fig:finite_size} we compare the contribution of the finite-size corrections of the critical temperature in 1D \eqref{Eq:bc1DFS} with the corrections for the near 2D and 2D critical temperature \eqref{Eq:bc2D1FS} and \eqref{Eq:bc2DFS}, respectively. We see directly, that in two dimensions the finite-size corrections are one order of magnitude smaller than in the 1D case. Thus, we deduce that when performing the crossover from 2D to 1D the importance of the finiteness of the system increases, as the finite-size corrections increase. 

\subsection{Condensate Fraction}
\label{ssec:condfrac}
In this section we calculate the condensate fraction $N_0/N$, where $N_0$ is the ground-state particle number, in the deep condensate limit, i.e.~for $N\gg N_\text c$. Thus, using Eq.~\eqref{Eq:NBEC} we have
\begin{align}\label{Eq:CondFrac}
    \frac{N_0}{N} \approx 1-\frac{N_\text c}{N}.
\end{align}
In 1D we find for the fraction $N_\text c /N$ by using the critical particle number \eqref{Eq:Nc1D}
\begin{align}\label{Eq:CondFrac1D}
	\left(\frac{N_\text c}{N}\right)_\text{1D} = \frac{T}{T_\text c}\left[1-\frac{\ln\left(T/T_\text c\right)}{\gamma-\ln(\hbar\Omega/k_\text B T_\text c)}\right],
\end{align}
so we have in leading order a linear temperature dependence.
\begin{figure}
    \centering
    \includegraphics[width=.75\linewidth]{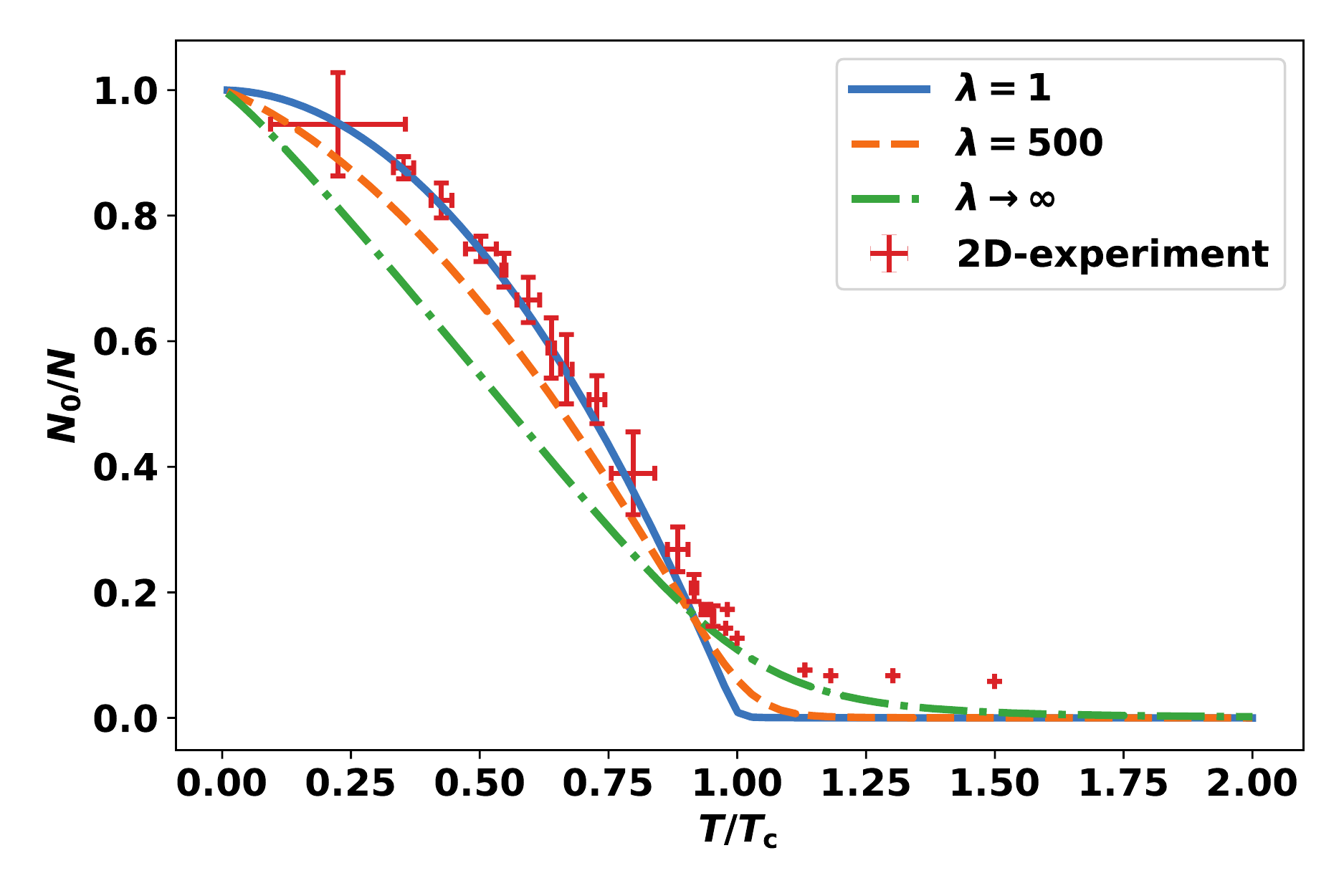}
    \caption{Condensate fraction $N_0/N$ for fixed particle number $N=100,000$. The blue (solid) line represents the isotropic 2D case, i.e.~$\lambda=1$, the orange (dashed) line is for $\lambda=500$, and the green (dash-dotted) line shows the 1D limit $\lambda\rightarrow\infty$. The red crosses are experimental values with the corresponding errors for the 2D case \cite{Liang}.}
    \label{fig:fraction}
\end{figure}
In contrast to this, when we approach two spatial dimensions, using the corresponding expression \eqref{Eq:Nc2D1}, we find in the leading order a quadratic dependence on the temperature
\begin{align}\label{Eq:CondFrac2D1}
	&\left(\frac{N_\text c}{N}\right)_{\approx\text{2D}} = \left(\frac{T}{T_\text c}\right)^2+\frac{1}{2\zeta(2)}\sqrt{\frac{g\zeta(2)}{N}}\left\{\left[\frac{T}{T_\text c}-\left(\frac{T}{T_\text c} \right)^2\right]\right.\\
	&\left.\times \left[-\frac{1}{2}\ln\frac{g\zeta(2)}{N}+\frac{\ln\Gamma(\lambda)-\ln\left(2\pi\sqrt{g\zeta(2)/N}\right)}{\lambda}-\psi_0(\lambda)+2\gamma\right]+\frac{T}{T_\text c}\left(1+\frac{1}{\lambda}\right)\ln\frac{T}{T_\text c}\right\}.\nonumber
\end{align}
In the two-dimensional limit Eq.~\eqref{Eq:CondFrac2D1} reduces to
\begin{align}\label{Eq:CondFrac2D}
	\left(\frac{N_\text c}{N}\right)_\text{2D} =& \left(\frac{T}{T_\text c}\right)^2+\sqrt{\frac{g\zeta(2)}{N}}\left\{\left[\frac{T}{T_\text c}-\left(\frac{T}{T_\text c} \right)^2\right]\frac{3\gamma-\ln(2\pi g\zeta(2)/N)}{2\zeta(2)}-\frac{T\ln(T/T_\text c)}{T_\text c\zeta(2)}\right\}. 
\end{align}
Figure \ref{fig:fraction} shows a numerical calculation of the temperature dependence of the condensate fraction for an experimentally realistic number of $N=100,000$ photons for different values of the trap-aspect ratio $\lambda$. The numerical calculation of the condensate fraction is done as follows. At first, we invert the particle number equation \eqref{Eq:N} in order to extract the dimensionless chemical potential $\tilde\mu$. We then use this value to calculate the ground-state population $N_0$ and, thus, the condensate fraction. We note that the isotropic 2D curve is in good agreement with the experiment of Ref.~\cite{Liang}. The discrepancy in the thermal phase is attributed to the finite resolution of the experimental apparatus. Moreover, we observe the inverted parabolic temperature dependence \eqref{Eq:CondFrac2D}. Also in the 1D case the curve agrees with the linear temperature dependence predicted in \eqref{Eq:CondFrac1D}. For the curve with an intermediate trap-aspect ratio of $\lambda=500$ the curve shows characteristics of both the 1D curve and as the temperature increases also of the 2D curve, meaning that here the effective dimension of the system changes from 1D to 2D.

To conclude this discussion, we remark that in the experimental situation of photon BECs the temperature is always fixed to the room temperature $T_0$, but the usual way for measuring thermodynamic quantities is to change the temperature $T$. So far this problem has been experimentally circumvented by varying the particle number and using that for integer dimensions $D=1,2$ an analytic correspondence is available in the form $T/T_\text c = (N_\text c/N)^{1/D}$ \cite{Damm2016}. Thus, we see already here, that this procedure is not well suited for a dimensional crossover, where one deals with non-integer dimensions. The second problem is that due to the finite system size also the thermodynamic quantities change with the particle number $N$. Instead of this, we propose to change the parameter $b$ via the trapping frequency $\Omega$. This is possible, since from its very definition we have $b_\text c/b = T/T_\text c$ and also $b_\text c/b = \Omega_\text c/\Omega$.\\
Finally, we estimate the experimentally achievable regions of $T/T_\text c$ in photon BECs for realistic total particle numbers of $N_\text{min} \sim 10^2$ up to $N_\text{max}\sim 10^5$ in both 2D and 1D. We find in 2D that the order of magnitude of the fraction $T/T_\text c$ ranges from 0.1 up to 10, whereas in 1D we have the range from 0.1 to 1. As all calculated quantities depend smoothly on the trap-aspect ratio, we expect that for a certain value of the trap-aspect ratio the reachable values of $T/T_\text c$ lie in between the corresponding 2D and 1D ratio, such that the phase transition is always observable.

\section{Specific Heat\label{Sec:Spec}}
In the following we calculate the specific heat $C_N$ for a constant particle number $N$ from the internal energy $U=\Pi+TS+\mu N$ \cite{Schwabl2006}, where $S$ denotes the entropy, according to
\begin{align}
    C_N = \frac{\partial U}{\partial T} + \frac{\partial U}{\partial\mu}\left(\frac{\partial \mu}{\partial T}\right)_N,
\end{align}
where the second term takes the condition of a fixed particle number into account. However, due to the complexity of the resulting formula, we restrict the discussion to the deeply condensed case at first. In this approximation we find for the internal energy
\begin{align}\label{Eq:U}
    U\approx & E_0(N_0+N_\text{c})+\hbar\Omega\lambda \Delta N+g\frac{\hbar\Omega}{b}\left[\frac{f'(\lambda)}{2}-\frac{f(\lambda)}{\lambda}\right]+ g\hbar\Omega\left[\frac{\zeta(2)}{b^2} -\frac{1}{2b}+ \frac{2}{\lambda b^3}\zeta_3\left(e^{-\lambda b}\right)\right.\nonumber\\
    &\left.+\frac{\lambda+1}{2\lambda b^2}\zeta_2\left(e^{-\lambda b}\right) + \frac{1}{24}\zeta_0\left(e^{-\lambda b}\right)\right],
\end{align}
where $\Delta N = N-N_\text{1D}$.
We note, that in this limit we have $\partial\mu/\partial T\approx0$, so the specific heat in the leading order of $1/b$ reads
\begin{align}\label{Eq:cN}
    C_N\approx g k_\text B\left[\frac{2\zeta(2)}{b} + \frac{6}{\lambda b^2}\zeta_3\left(e^{-\lambda b}\right) + \frac{3\lambda+1}{b}\zeta_2\left(e^{-\lambda b}\right)\right].
\end{align}
Therefore, taking into account Eq.~\eqref{Eq:bc1D}, in the one-dimensional limit the specific heat is given by
\begin{align}\label{Eq:c1D}
    C_{N, \text{1D}} = Nk_\text B \frac{2\zeta(2)}{\gamma-\ln(\hbar\Omega/k_\text B T_\text c)}\frac{T}{T_\text c},
\end{align}
and in leading order in $b$ near 2D we obtain from Eq.~\eqref{Eq:bc2D1LT} together with Eq.~\eqref{Eq:bc2D1FS}
\begin{align}\label{Eq:c2D1}
    C_{N, \approx\text{2D}} \approx& 6\lambda Nk_\text B \left(\frac{T}{T_\text c}\right)^2\frac{\zeta(3)}{\zeta(2)}\left\{1 - \frac{1}{2\zeta(2)}\sqrt{\frac{g\zeta(2)}{\lambda N}}\right.\nonumber\\
    &\left.\times\left[2\gamma-\psi_0(\lambda)-\frac{\ln\Gamma(\lambda)-\ln\left(2\pi\sqrt{g\zeta(2)/\lambda N}\right)}{\lambda}-\frac{1}{2}\ln\frac{g\zeta(2)}{\lambda N}\right]\right\}.
\end{align}
In two dimensions Eq.~\eqref{Eq:c2D1} reduces to
\begin{align}\label{Eq:c2D}
    C_{N, \text{2D}} \approx& 6Nk_\text B \left(\frac{T}{T_\text c}\right)^2\frac{\zeta(3)}{\zeta(2)}\left\{1 - \frac{1}{2\zeta(2)}\sqrt{\frac{g\zeta(2)}{N}}\left[3\gamma-\ln\left(2\pi\frac{g\zeta(2)}{N}\right)\right]\right\}.
\end{align}
After deriving these analytic formulas in the deeply condensed case, we discuss now the obtained results and compare them with the experimental results of Ref.~\cite{Damm2016}.
\begin{figure}
    \centering
    \includegraphics[width=.7\linewidth]{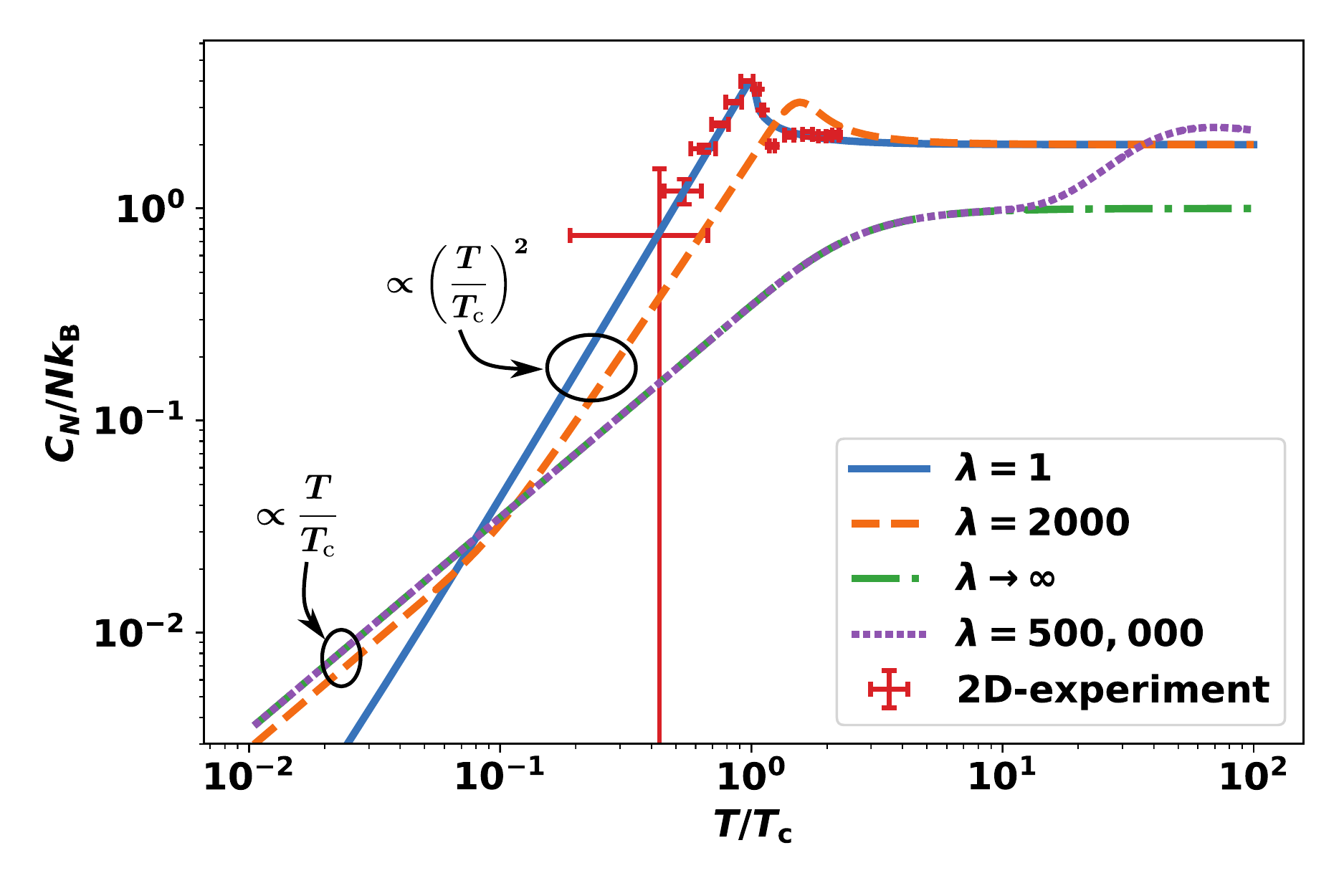}
    \caption{Temperature dependence of specific heat $C_N$ for the particle number $N=100,000$. The blue (solid) line covers the isotropic 2D case, i.e.~$\lambda=1$, the orange (dashed) line is for $\lambda=2000$, the pink (dotted) line represents $\lambda=500,000$, and the green (dash-dotted) line is for $\lambda\rightarrow\infty$. The red crosses are experimental values with the corresponding errors for the 2D case from Ref.~\cite{Damm2016}.}
    \label{fig:cN}
\end{figure}
In Fig.~\ref{fig:cN} we plot the full specific heat for different values of the trap-aspect ratio $\lambda$ as a function of the temperature $T$. In the two-dimensional case we find the expected $\lambda$-like transition with the high-temperature limit of $C_N = 2 Nk_\text B$, where the latter is in accordance with the Dulong-Petit law. Note, that in our case of a finite system the specific heat does not undergo a jump at the critical point as it occurs in the thermodynamic limit \cite{Klunder2009}. Instead, it remains a continuous function. Furthermore, we point out that in 1D the specific heat is always a continuous function at the phase transition which explicitly includes the thermodynamic limit, cf.~\cite{Klunder2009}. However, as we increase the trap-aspect ratio, we see that this characteristic $\lambda$-shaped behaviour near the critical temperature $T_\text c$ gradually washes out and a plateau emerges just above the critical temperature. This plateau has the value $C_N = Nk_\text B$ and, thus, resembles the one-dimensional Dulong-Petit law, meaning that here the system, indeed, behaves as one-dimensional. By further increasing the temperature the system approaches again the 2D Dulong-Petit law. However, the dotted line in Fig.~\ref{fig:cN} is slightly above $C_N = N k_\text B$ since here the system undergoes the crossover from 2D to 1D producing the corresponding $\lambda$-like form near the crossover temperature. The reason for the described behaviour is as follows. The temperature can be seen as a measure for which states can be occupied, namely as the temperature raises also states with higher energies are populated. Therefore, we can invert the 1D condition \eqref{Eq:Crit_1D} and define for a fixed trap-aspect ratio $\lambda$ the effective 1D temperature $T_\text{1D} = \lambda\hbar\Omega/k_\text B$. We note, that the system is in the 1D regime if $T<T_\text{1D}$, as here the occupation in the excited states in the squeezed direction is exponentially suppressed, and otherwise in the 2D regime, since then the thermal energy is large enough to have also states in the squeezed direction populated.\\
A similar behaviour is well known in the literature, e.g.~for the thermodynamics of molecular gases \cite{Schwabl2006}. At low temperatures only the translational degrees of freedom of the molecules can be thermally excited and, thus, only those can contribute to the specific heat. Increasing the temperature above a certain threshold allows the molecules to rotate such that these degrees of freedom additionally contribute to the specific heat. Increasing the temperature even further allows also the vibrational modes of the molecules to be thermally excited.\\
We note the different behaviour of the specific heat in the low-temperature limit, which is worked out in Eqs.~\eqref{Eq:c1D} and \eqref{Eq:c2D}. Thus, in contrast to the condensate fraction and the critical particle number, using the specific heat can be instrumental to define and to determine the effective dimension of the system both in the low and the high temperature limit.

\section{Phase Diagram and Effective Dimension \label{Sec:Phase_Dim}}
\begin{figure}
    \centering
    \includegraphics[width=\linewidth]{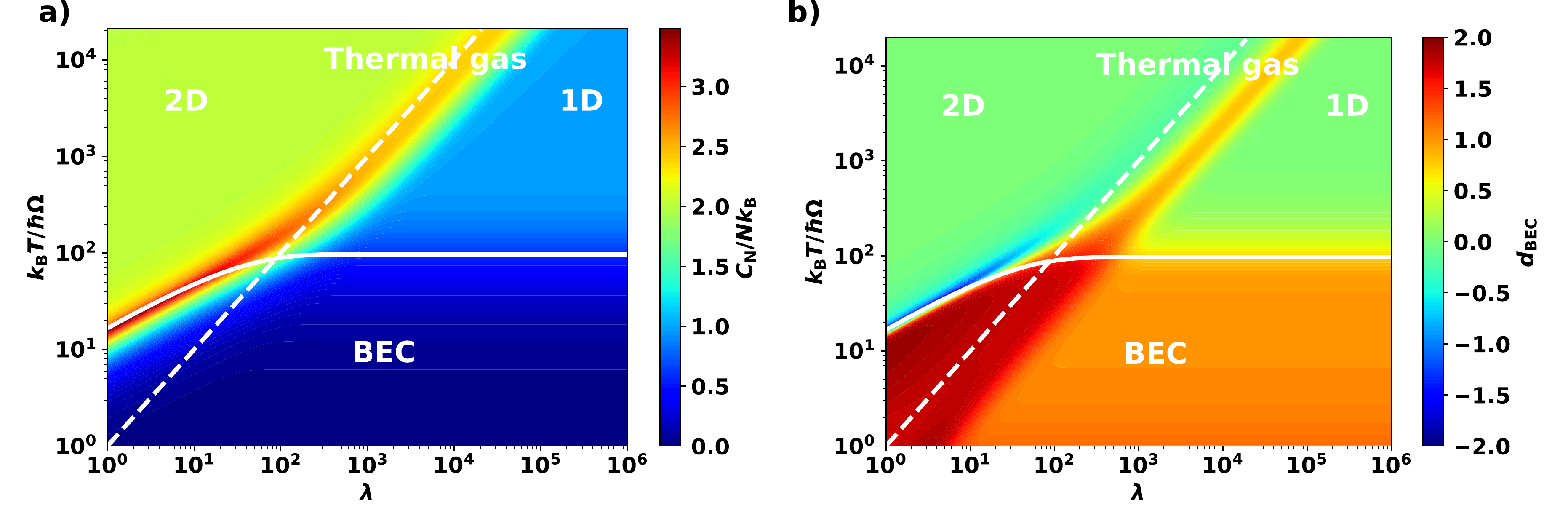}
    \caption{\textbf{a)} Phase diagram of the ideal Bose gas at the dimensional crossover by plotting colour coded the specific heat \eqref{Eq:cN} as a function of both the temperature and the trap-aspect ratio. \textbf{b)} Effective dimension of the BEC phase according to the definition \eqref{Eq:dimBEC}. In both plots the solid white line shows the critical temperature following from inverting \eqref{Eq:Nc}, whereas the dashed white curve depicts the criterion \eqref{Eq:Crit_1D} of being quasi 1D. Both calculations have been performed for $N=1000$ particles.}
    \label{fig:phase}
\end{figure}
Finally, we analyse how the phase diagram of the system changes as a function of the trap aspect ratio $\lambda$ and the temperature $T$ by plotting the specific heat $C_N$ in Fig.~\ref{fig:phase} a). However, due to numerical reasons we are forced to use a small particle number of $N=1000$ photons, as otherwise we are not able to cover such a large range of parameters $\lambda$ and $k_\text B T/\hbar\Omega$.
\begin{figure}
    \centering
    \includegraphics[width=\linewidth]{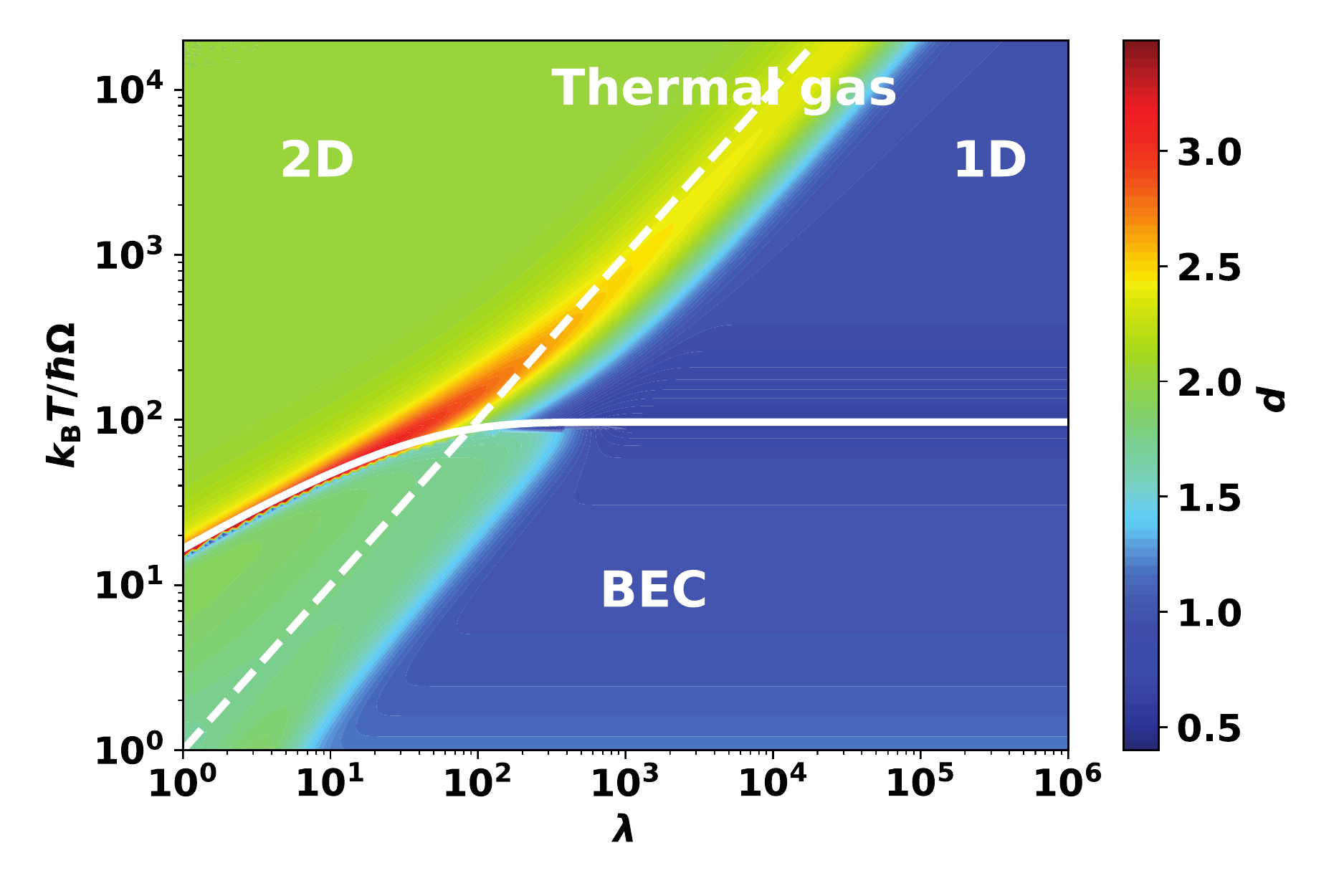}
    \caption{Effective dimension $d$ of the ideal Bose gas at the dimensional crossover as defined by \eqref{Eq:dim} as a function of both the temperature and the trap-aspect ratio. The solid white curve shows the critical temperature obtained by inverting \eqref{Eq:Nc}, whereas the dashed white line indicates the quasi-1D criterion \eqref{Eq:Crit_1D}. The calculation has been performed for $N=1000$ particles.}
    \label{fig:dimension}
\end{figure}
At first we note that the phase transition from the BEC to the thermal phase happens at the critical temperature $T_\text c$, which is calculated by inverting the critical photon number \eqref{Eq:Nc} with the limiting cases \eqref{Eq:bc1D} and \eqref{Eq:bc2DLT}. Moreover, in the thermal phase, we can directly read off the effective dimension of the system according to the respective Dulong-Petit law, as is explained at the end of Sec.~\ref{Sec:Spec}. The dashed white line depicting the lower bound $\lambda_\text{1D} = k_\text B T/\hbar\Omega$ of criterion \eqref{Eq:Crit_1D}, discriminates between the different dimensional behaviour.
However, therefrom we can only learn about the effective dimension in the thermal phase. From Eqs.~\eqref{Eq:c1D} and \eqref{Eq:c2D}, though, we read off, that in the condensed regime the effective dimension follows from the polynomial dependency of the specific heat on the temperature. Therefore, we suggest to define as the effective dimension in the BEC phase the double-logarithmic derivative
\begin{align}\label{Eq:dimBEC}
    d_\text{BEC} = -\frac{1}{Nk_\text B}\frac{\partial \ln C_N}{\partial \ln b}.
\end{align}
Figure \ref{fig:phase} b) shows the corresponding results as a function of both the temperature $T$ and the trap aspect ratio $\lambda$. We note that in the thermal phase this definition yields a constant value of 0 due to the Dulong-Petit law except right at the crossover from 2D to 1D. Thus, this definition cannot be used in the thermal case to determine the effective system dimension. In the BEC phase, however, we find values between 1 and 2 according to the limiting cases deduced from Eqs.~\eqref{Eq:c1D} and \eqref{Eq:c2D}. Near the phase transition, however, this definition fails as here the quantitative behaviour of the specific heat changes yielding e.g.~negative values of $d_\text{BEC}$. The precise value of the BEC dimension is determined by the temperature and trap-aspect ratio. We see that, for an increasing trap-aspect ratio, the system behaves, indeed, quasi one-dimensional.\\
Summarising the two observations from Fig.~\ref{fig:phase}, we suggest to define the effective dimension of the system by
\begin{align} \label{Eq:dim}
    d = \begin{cases} 
    C_N /(Nk_\text B), &\text{in the thermal phase},\\
    d_\text{BEC},&\text{in the BEC phase},
    \end{cases}
\end{align}
where $d_\text{BEC}$ is defined in \eqref{Eq:dimBEC}.
With this we are able to describe the effective dimension of the system in both the Bose-condensed and the thermal regime. In Fig.~\ref{fig:dimension} we plot the definition \eqref{Eq:dim} in dependence of both the temperature and the trap aspect ratio. However, as Fig.~\ref{fig:dimension} shows, the definition \eqref{Eq:dimBEC} yields a non-continuous effective dimension at the phase boundaries as here the slopes of the specific heat change. This can be read off from the reddish area in the plot. Nevertheless, we also note, that the effective dimension of the system changes from 2D to effective 1D in agreement with the criterion \eqref{Eq:Crit_1D}. We remark, that in the crossover region both the temperature and the trap-aspect ratio determine the effective dimension of the system.
Finally, we point out that trap-aspect ratios up to $\lambda\sim10^3$ are experimentally realisable, which is due to the expected resolution of the mirror fabrication method \cite{PrivFrank}. Consequently, according to Fig.~\ref{fig:dimension} the onset of the effective 1D region is reachable at room temperature, where we have $k_\text B T_0/\hbar \Omega\approx 160$.

\section{Conclusions}
In this paper we present an analytical description of the dimensional crossover from 1D to 2D for an ideal Bose gas in terms of a dimensional expansion, see Eq.~\eqref{Eq:Gen_pontential}. We find the same structure for all investigated thermodynamic quantities, such as the critical particle number, the condensate fraction, and the specific heat, namely that the 1D expression gets corrected by terms yielding the 2D result. Furthermore, from the specific heat we are able to define an effective dimension $d$, given by Eq.~\eqref{Eq:dim}, in both the BEC and the thermal phase. This definition shows a change of the effective dimension, which is consistent with the criterion \eqref{Eq:Crit_1D}. But we also note, that this definition has a minor drawback as it produces a non-continuous behaviour of the effective dimension near the phase boundary, as can be seen in Fig.~\ref{fig:dimension}. However, our results allow to determine the effective dimension of the system for a given temperature and trap-aspect ratio. We especially focus on how to determine the effective dimension by examining the polynomial dependency of the specific heat in the BEC case and by observing the Dulong-Petit law in the thermal regime. We remark, that our calculational approach, which is based on an expansion in the smallness parameter $\hbar\Omega/(k_\text B T)$, is especially suitable for photon gases, where this value is of the order of a few per mille.\\
The present work could be extended to also determine the spatio-temporal behaviour of the correlation function of the ideal Bose gas at the dimensional crossover, which has already been measured for an isotropic two-dimensional photon gas \cite{Damm2017}. Concerning the fact, that in the 2D photon BEC experiments a retarded thermo-optic interaction is dominant, despite of an additional negligible contact interaction \cite{Klaers2010a, Stein2019, Radonjic2018}, it is an interesting question, whether this is still true in the quasi-1D case. Moreover, for a more realistic modelling of the experiments, one needs to include also the pump and the decay processes, as a photon gas in a dye-filled microcavity is intrinsically an open system. A recent study \cite{Ozturk2021} indeed shows that due to the open-dissipative character of the system the higher correlation function shows a phase transition, which does not exist in closed systems. In a second attempt one should also include the effective photon-photon interaction, as it is known, that e.g.~contact interactions increase along a dimensional crossover. 
Another research direction would be to investigate in view of the dimensional crossover different potential landscapes, such as potentials with arbitrary exponents, c.f.~\cite{Bagnato1991}, or even anharmonic potentials \cite{Balaz2010}.

\ack
We thank Antun Bala\v{z}, Erik Busley, Georg von Freymann, Milan Radonji\'{c}, Julian Schulz, Kirankumar Karkihalli Umesh and Frank Vewinger for insightful discussions. E. S. and A. P. acknowledge financial support by the Deutsche Forschungsgemeinschaft (DFG, German Research Foundation) via the Collaborative Research Center SFB/TR185 (Project No. 277625399).

\section*{References}
\bibliographystyle{unsrt}
\bibliography{lib}

\newpage
\appendix
\section{Cutoff Regularisation \label{App:Regularisation}}
The aim of this appendix is to work out the behaviour of the auxiliary functions $I(a,b,l)$ defined in Eq.~\eqref{Def:I} for integer $l$ and $a,b>0$ and also to provide a procedure allowing to approximate these functions analytically. First we start with two recursion relations obeyed by the auxiliary functions. For increasing the integer $l$ we have
\begin{align}\label{Eq:Rec1}
	I(a, b, l+1) = -\frac{1}{b}\frac{\partial}{\partial a} I(a, b, l),
\end{align}
whereas decreasing $l$ yields correspondingly
\begin{align}\label{Eq:Rec2}
	I(a, b, l-1) = b \int_a^\infty dx~I(x,b, l).
\end{align}
Thus, from the analytical knowledge of one particular $I(a, b, l^*)$ all other functions $I(a, b, l)$ can be calculated analytically.

\subsection{Special Case $l^*=0$}
It turns out, that the case $l^* = 0$ can be calculated analytically for small values of $b$. 
According to the definition \eqref{Def:I} we start with
\begin{align}\label{Eq:AppI}
	I(a,b,0) = \sum_{k=1}^\infty \frac{e^{-abk}}{1-e^{-bk}}.
\end{align}
In order to calculate expression \eqref{Eq:AppI}, we follow Ref.~\cite{Klunder2009} and perform an expansion for small values of $b$. However, the first step is to include also the $k=0$ term in the summation \eqref{Eq:AppI}. As this is a divergent term, we add and subtract the first three terms of the corresponding Laurent series. Note, that in Ref.~\cite{Klunder2009} only the first term of the Laurent series is introduced yielding an approximation up to $\mathcal{O}(b^0)$. However, here we need higher order terms for obtaining a converging result for the two-dimensional case. Thus, we have 
\begin{align}\label{Eq:AppI1}
	I(a,b,0) = \sum_{k=0}^\infty e^{-abk}\left(\frac{1}{1-e^{-bk}}-\frac{1}{bk}-\frac{1}{2}-\frac{bk}{12}\right)
	+\sum_{k=1}^\infty e^{-abk} \left(\frac{1}{bk} + \frac{1}{2} + \frac{bk}{12}\right)+\mathcal{O}(b^2).	
\end{align}
In the first term we replace the summation by an integral using the Euler-Maclaurin formula for a smooth function $f(n)$
\begin{align}\label{Eq:Euler}
    \sum_{n=0}^\infty f(n)\approx \int_0^\infty dn~f(n) + \frac{1}{2}\left[f(0)+f(\infty)\right].
\end{align}
Due to the construction of expression \eqref{Eq:AppI1}, all higher terms in the Euler-Maclaurin series \eqref{Eq:Euler} vanish exactly.
In the second term we recognise the polylogarithmic functions $\zeta_n(x)$ with $n=-1,0,+1$. Thus, we have
\begin{align}
	I(a,b,0) = &\int_{0}^\infty dk~ e^{-abk}\left(\frac{1}{1-e^{-bk}}-\frac{1}{bk}-\frac{1}{2}-\frac{bk}{12}\right)\nonumber\\
	&+\frac{1}{b}\zeta_1\left(e^{-ab}\right) + \frac{1}{2}\zeta_0\left(e^{-ab}\right) + \frac{b}{12}\zeta_{-1}\left(e^{-ab}\right)+\mathcal{O}(b^2).		
\end{align}
Whereas in Ref.~\cite{Klunder2009} the remaining integrals are solved by using a dimensional regularisation, we introduce here an infrared cutoff $\Lambda$ as the integrands are divergent for $k\rightarrow0$:
\begin{align}\label{Eq:I}
	I(a,b,0) = &\lim_{\Lambda\rightarrow0} \int_{\Lambda}^\infty dk~ e^{-abk}\left(\frac{1}{1-e^{-bk}}-\frac{1}{bk}-\frac{1}{2}-\frac{bk}{12}\right)\nonumber\\
	&+\frac{1}{b}\zeta_1\left(e^{-ab}\right) + \frac{1}{2}\zeta_0\left(e^{-ab}\right) + \frac{b}{12}\zeta_{-1}\left(e^{-ab}\right)+\mathcal{O}(b^2).		
\end{align}
First, we obtain
\begin{align}
	\int_{\Lambda}^\infty dk~(bk)^ne^{-abk} = \frac{1}{a^{n+1}b}\Gamma(n+1, ab\Lambda),
\end{align}
where $\Gamma(s,x)$ is the upper incomplete $\Gamma$ function. For small $\Lambda$ we find in leading order
\begin{align}\label{Eq:Gamma1}
	\Gamma(0, ab\Lambda) \approx \frac{-\gamma-\ln(ab\Lambda)}{b},
\end{align}
whereas the incomplete $\Gamma$ functions with indices $n\geq1$ simply reduce to the standard $\Gamma$ functions:
\begin{align}\label{Eq:Gamma2}
    \Gamma(n, ab\Lambda) \approx \Gamma(n),~n\geq1.
\end{align}
In the remaining first integral of Eq.~\eqref{Eq:I} we substitute $x = e^{-bk}$ and calculate by using the incomplete beta function,
\begin{align}
	B(x;a,b) = \int_0^x dt~t^{a-1}(1-t)^{b-1},		
\end{align}
the integral
\begin{align}\label{Eq:Beta}
	\int_{\Lambda}^\infty dk~ \frac{e^{-abk}}{1-e^{-bk}} = \frac{1}{b}B(e^{b\Lambda};a,0).
\end{align}
This yields in the limit of small $\Lambda$ 
\begin{align}\label{Eq:Beta1}
	B(e^{-b\Lambda};a,0) \approx-\ln(b\Lambda)-\gamma-\psi_0(a).
\end{align}
Inserting Eqs.~\eqref{Eq:Gamma1}, \eqref{Eq:Gamma2} and \eqref{Eq:Beta1} into Eq.~\eqref{Eq:I} we finally have
\begin{align}\label{Eq:I0}
	I(a, b,0) =  &\frac{1}{b}\left[\ln(a)-\psi_0(a)-\frac{1}{2a}-\frac{1}{12a^2}\right]+\frac{1}{b}\zeta_1\left(e^{-ab}\right)+\frac{1}{2}\zeta_0\left(e^{-ab}\right)+\frac{b}{12}\zeta_{-1}\left(e^{-ab}\right)\nonumber\\&+\mathcal{O}(b^2),
\end{align}
which coincides with the result in Ref.~\cite{Klunder2009}, apart from the additional higher order terms. 
In the following we calculate $I(a,b,-1)$ from applying the recurrence relation \eqref{Eq:Rec2}. To this end we use the Stirling formula \cite{Gradshteyn2007},
\begin{align}
\ln\Gamma(z) \approx z(\ln z-1)-\frac{1}{2}\ln(2\pi z),
\end{align}
for regularising the upper integration limit in Eq.~\eqref{Eq:Rec2}, and obtain
\begin{align}\label{Eq:IM1}
	I(a, b, -1) = f(a) +\frac{1}{b}\zeta_2\left(e^{-ab}\right)+\frac{1}{2}\zeta_1\left(e^{-ab}\right)+\frac{b}{12}\zeta_{0}\left(e^{-ab}\right)+\mathcal{O}(b^2),
\end{align}
where we defined
\begin{align}\label{Eq:DefI}
    f(a) = \frac{1}{2}\ln\left(\frac{a}{2\pi}\right) -a\left[\ln(a)-1\right]+\ln\Gamma(a) - \frac{1}{12a}.
\end{align}
This result is still correct to order $\mathcal{O}(b^2)$, because the recurrence relation \eqref{Eq:Rec2} preserves the corresponding order.

\subsection{Resummation for Second Dimension}
In \eqref{Eq:DPi} we have seen, that we also need to calculate a sum over the auxiliary functions \eqref{Def:I}.
With the result \eqref{Eq:IM1} we can also analytically approximate the sum
\begin{align}\label{Eq:ISum1}
	S=\sum_{n=1}^\infty I(a+\lambda n, b, -1)
\end{align}
by using again the Euler-Maclaurin series \eqref{Eq:Euler}. Thus, we obain the approximation
\begin{align}\label{Eq:EulerS}
    S\approx \int_1^\infty dn~I(a+\lambda n, b, -1) + \frac{1}{2}I(a+\lambda, b, -1).
\end{align}
Taking Eq.~\eqref{Eq:IM1} into account, we have
\begin{align} \label{Eq:AppS}
    S =
    &\frac{1}{\lambda b^2} \zeta_3\left(e^{-(a+\lambda)b}\right) + \frac{1}{2\lambda b}\zeta_2\left(e^{-(a+\lambda)b}\right)+ \frac{1}{12\lambda}\zeta_1\left(e^{-(a+\lambda)b}\right) 
    + \frac{1}{\lambda} \int_{a+\lambda}^\infty dy~f(y) \nonumber\\
    &+ \frac{1}{2}I(a+\lambda, b, -1)+\mathcal{O}(b^2).
\end{align}
We note that the error, stemming from the Euler-Maclaurin approximation in Eq.~\eqref{Eq:EulerS}, cannot be evaluated in a systematic way. However, we show in the next section, that the performed approximation yields errors, which are small in the relevant parameter regime of photon gases.

\subsection{Analytical vs. Numerical Summation}
\begin{figure}
    \centering
    \includegraphics[width=.75\linewidth]{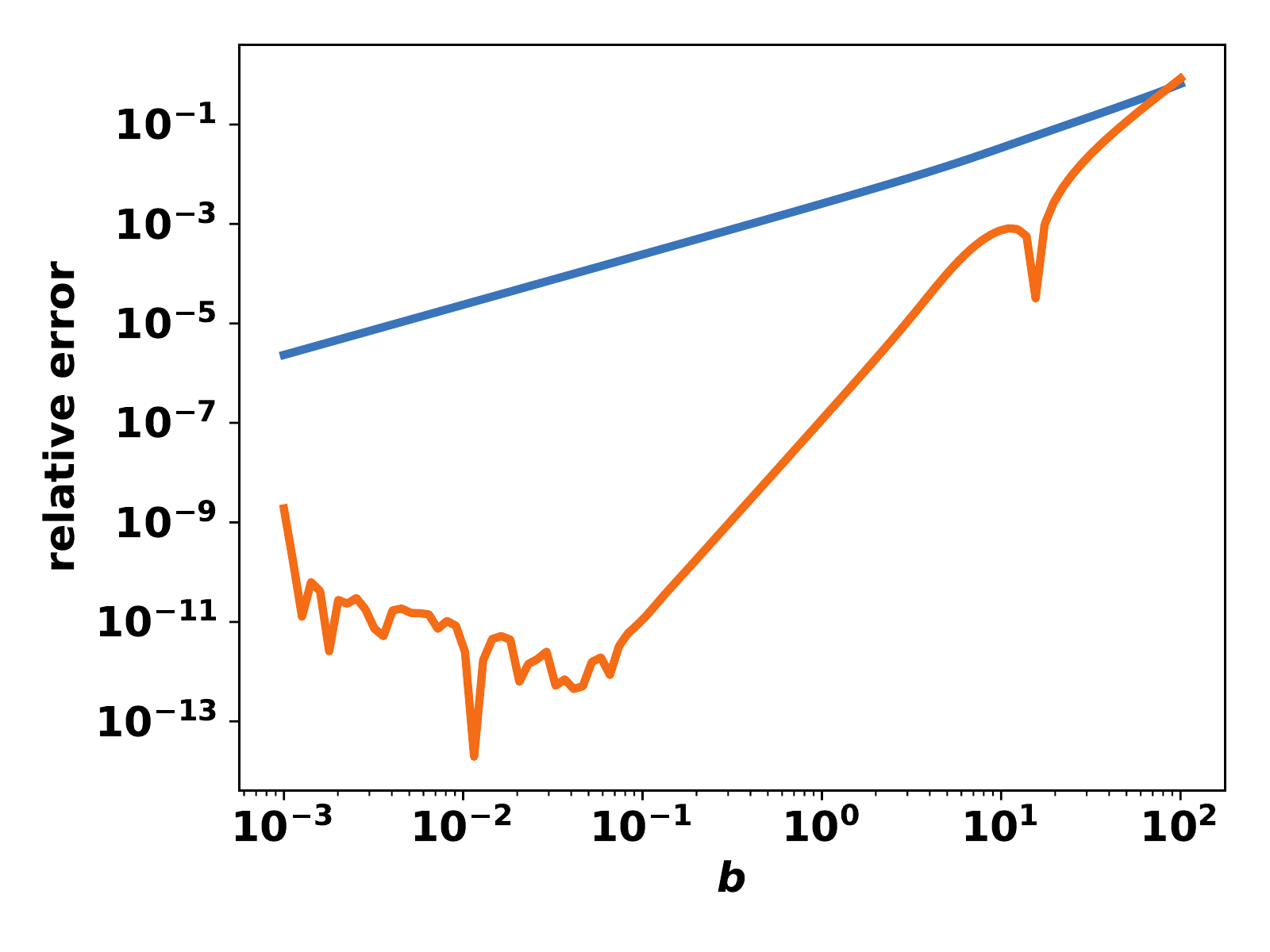}
    \caption{Relative error of analytical approximation of the one-dimensional sum \eqref{Eq:I0} with respect to the numerical evaluation of Eq.~\eqref{Eq:AppI} (orange line). The blue line shows the relative error by using the approximation performed in Ref.~\cite{Klunder2009}.}
    \label{fig:rel_1D}
\end{figure}
\begin{figure}
    \centering
    \includegraphics[width=.75\linewidth]{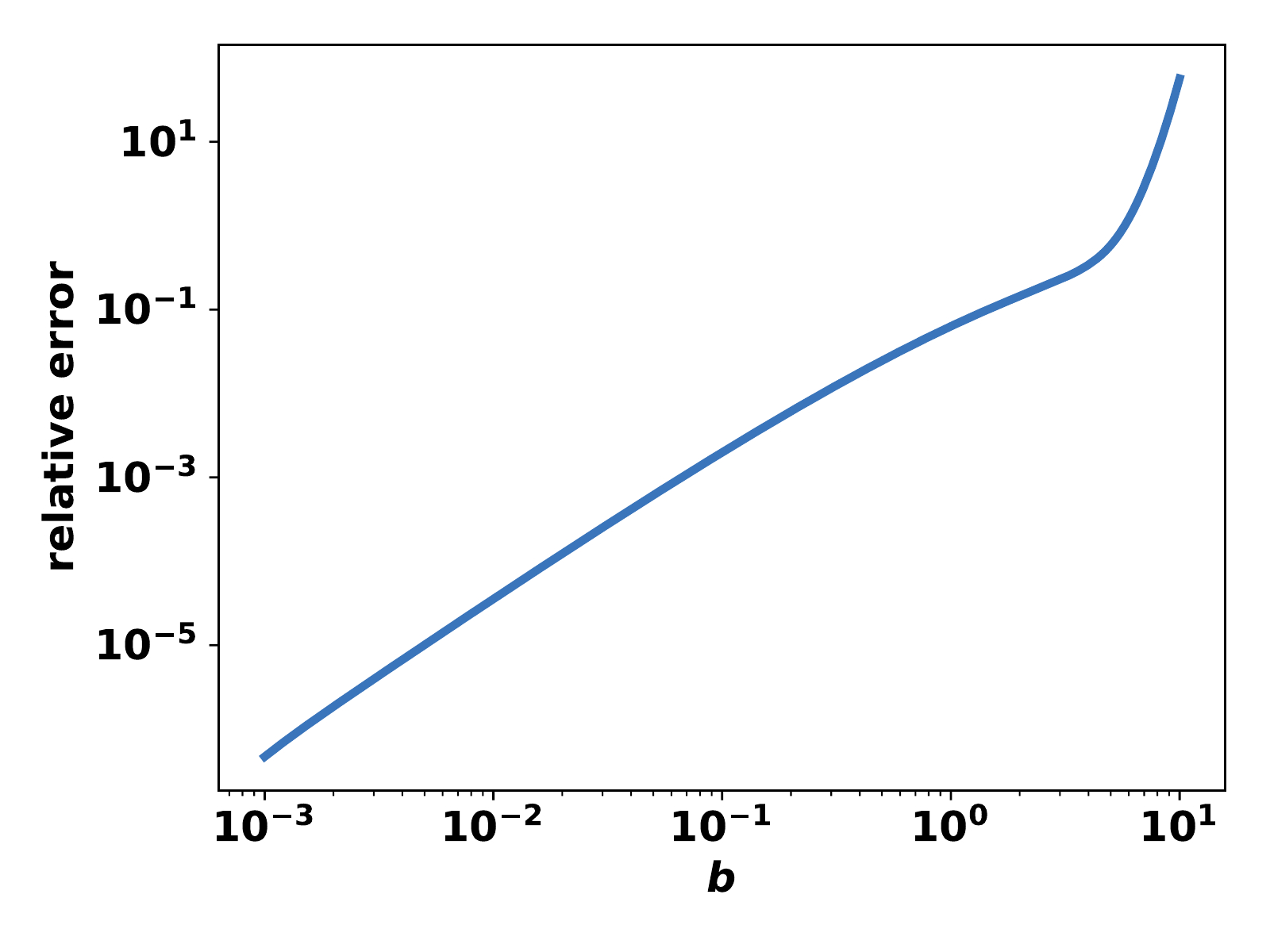}
    \caption{Relative error of analytical approximation of the correction terms leading to the second dimension \eqref{Eq:AppS} with respect to the numerical evaluation in Eq.~\eqref{Eq:AppS2D}.}
    \label{fig:rel_2D}
\end{figure}
Finally, we compare the analytical results from the preceding sections with a numerical summation of Eq.~\eqref{Eq:AppI} itself. Figure \ref{fig:rel_1D} shows the relative error of the numerical approximation \eqref{Eq:I0} with respect to the direct numerical evaluation of the sum \eqref{Eq:AppI}. The orange line shows our result, whereas the blue line shows the accuracy achieved in Ref.~\cite{Klunder2009}. At first, we note that both results yield a good approximation as $b$ tends to 0. However, as we use additional terms from the Laurent series in Eq.~\eqref{Eq:AppI1}, the accuracy of our result is increased compared to the result from Ref.~\cite{Klunder2009}.\\
In order to analyse the error of the 2D result we first note that the sum \eqref{Eq:ISum1} can also be performed by using the definition \eqref{Eq:AppI} and interchanging the summation signs, which yields
\begin{align}\label{Eq:AppS2D}
    S_\text{2D} = \sum_{k=1}^\infty \frac{e^{-abk}}{k(1-e^{-bk})(e^{\lambda bk}-1)}.
\end{align}
Note, that due to the factor $1/k$ this expression cannot be treated analytically along the philosophy of Ref.~\cite{Klunder2009} and this appendix. However, expression \eqref{Eq:AppS2D} can be used as a numerical comparison with the analytical approximation obtained in \eqref{Eq:AppS}.
The relative error of the approximation of the two-dimensional sum \eqref{Eq:ISum1} is shown in Fig.~\ref{fig:rel_2D}. It reveals, as suspected, the same overall behaviour as the approximation of the one-dimensional sum, namely that the approximation gets better at small $b$ and worse at large values of $b$.

\end{document}